\documentclass[aps,prd,fleqn,superscriptaddress]{revtex4}
\pdfoutput=1
\usepackage{graphicx,color,natbib}
\usepackage{amsmath,amssymb,amsfonts}

\usepackage{epsfig,epsf}
\usepackage{dcolumn}
\usepackage{float}
\usepackage{bm}

\newcommand{\bse}{\begin{subequations}}
\newcommand{\ese}{\end{subequations}}
\newcommand{\be}{\begin{equation}}
\newcommand{\ee}{\end{equation}}
\newcommand{\bea}{\begin{eqnarray}}
\newcommand{\eea}{\end{eqnarray}}
\newcommand{\ba}{\begin{array}}
\newcommand{\ea}{\end{array}}

\def\a{\alpha}

\newcommand{\mm}{(\frac{\mu}{T})_*}

\input amssym.def
\input amssym.tex

\usepackage[colorlinks=true, linkcolor=blue, bookmarks=true]{hyperref}
\begin{document}

\title{Complexity and uncomplexity during energy injection}

\author{Mahsa Lezgi\footnote{s\_lezgi@sbu.ac.ir}}
\affiliation{Department of Physics, Shahid Beheshti University, Tehran, Iran}
\author{Mohammad Ali-Akbari\footnote{m\_aliakbari@sbu.ac.ir}}
\affiliation{Department of Physics, Shahid Beheshti University, Tehran, Iran}

\begin{abstract}
We consider a strongly coupled field theory with a critical point and nonzero chemical potential at finite temperature, which is dual to an asymptotically AdS charged black hole. We study the evolution of the rescaled holographic subregion complexity near and far from the critical point. We explain two distinct concepts of complexity in this theory and discuss that the state under study is complex based on how much information is needed to specify the state and is simple according to how many operations have to be done to reach the state. It has been reported before that time evolution of holographic subregion complexity contradicts the second law of complexity in these AdS-Vaidya-like geometries, but we try to provide a compatible interpretation. We justify decreasing of complexity using an increasing number of microstates of the mixed state and speculate about the description of the relative complexity of the initial state and the final state as a resource. We propose that in this process complexity of the mixed state decreases and complexity of the environment increases. We also observe that in this model, the rescaled holographic subregion complexity is a good observable for probing the dynamical critical exponent. 
\end{abstract}
\maketitle
\tableofcontents
\section{Introduction}
The evolution of out-of-equilibrium systems towards equilibrium is one of the important areas of physics to study. If these systems are far-from-equilibrium and strongly coupled, they are much more complicated to investigate. Gauge-gravity duality or more generally the holographic idea, by mapping strongly coupled field theories to weakly coupled gravity, provides a useful framework to study different non-perturbative problems, including far-from-equilibrium phenomena \citep{CasalderreySolana:2011us,rangamani}. Quark-gluon plasma produced in heavy ion collisions at the Relativistic Heavy Ion Collider (RHIC) is a good example of such a strongly coupled and far-from-equilibrium medium. Experimental observations show the time scale that is needed for this plasma to reach thermal equilibrium is significantly shorter than one that is expected from perturbative methods \cite{one}. Holographic thermalization proposes that such a thermalization process is dual to the process of the formation of a black hole in the bulk, for instance the collapse of a thin-shell matter described by an AdS-Vaidya metric \cite{bala}. A far-from-equilibrium state in field theory can be prepared by injection of energy using an external source and time evolution of this far-from-equilibrium system can be probed by local and non-local observables.

According to the holographic idea, a connection has been developed between quantities in quantum information theory and certain geometric quantities in the bulk geometry. For example, the Hubney-Ryu-Takayanagi proposal for entanglement entropy as a measure of quantum correlation of a pure quantum state, is a simple geometric prescription that passes many tests successfully \cite{Nishioka:2009un,Takayanagi}. Holographic entanglement entropy can be used to classify the various phase transitions and critical points \cite{kleb,kleb2}.

Complexity is another main concept in quantum information theory. Complexity of a state is defined as the minimum number of simple gates needed to produce it from some reference state \cite{john}. In the context of quantum field theory, complexity refers to the minimum number of unitary operators needed to prepare a target state from a reference state \cite{circuit}. In other words, complexity can classify various quantum states based on the difficulty of their creation. There exists two holographic prescriptions, namely the CV (complexity=volume) conjecture and the CA (complexity=Action) conjecture \cite{susskind1,susskind2}. These proposals have been introduced for the complexity of a pure state in the whole boundary system and they both can extend to be defined on subregions corresponding to the complexity of mixed states \cite{comments,alishahiha}. Inspired by the Hubney-Ryu-Takayanagi proposal, the complexity for a subsystem on the boundary is equal to the volume of codimensional-one hypersurface enclosed by Hubney-Ryu-Takayanagi surface, which is known as holographic subregion complexity (HSC) \cite{alishahiha}. Some works on CV and CA conjecture and HSC for various gravity models can be found in \cite{volume,zhang1,zhang2,renormalization,faregh,
mozaffar,mozaffar2,sub,asadi,mahsa}.

Complexity is like the entropy of an auxiliary classical system, which describes the evolution of operator $U(t)$ on $SU(2^{k})$ \cite{uncom1}. This description opens the window to a topic known as thermodynamics of complexity. For example, the second law of complexity is just the usual second law of thermodynamics applied to the auxiliary classical system. Moreover, in the thermodynamical view the lack of entropy or the difference between the maximum entropy and the actual entropy i.e. {\it{negentropy}}, is a resource for doing a work. There exists a concept analogous to negentropy in the thermodynamics of complexity. This parallel definition of the resource for complexity, called {\it{uncomplexity}} is the difference between maximum possible complexity and the actual complexity. In this case, work refers to doing directed computation, meaning a computation with a goal which is called {\it{computational work}} \cite{uncom1,uncom2}.

In this paper we investigate the evolution of HSC on a far-from-equilibrium state, dual to a Vaidya charged black hole, in the presence of temperature and chemical potential. We consider a gravity background dual to a field theory with a critical point and study the behavior of the relaxation time and HSC as the system moves towards the critical point. The questions we are interested in are: is HSC a proper observable to probe the dynamical critical exponent? Does the existence of a critical point in the theory help to distinguish between two concepts of complexity for determining whether a state is complex or simple? Considering a black hole as a computational machine, can a resource be defined for it in this time evolution? We start by introducing the background and then compute the HSC on the background and discuss its behavior to answer above questions.

\section{Review on the background}
A critical point can be described as a point at which the line of the first-order phase transition must end. Field theory observables can be studied near the critical point and one can also compute the associated (dynamical) critical exponent. Therefore, here we review a charged black hole solution which is holographically dual to a strongly coupled field theory with a critical point. 
The metric we are interested in, called charged dilaton AdS \cite{thermalization}, is
\begin{equation}
ds^2=-N(z)f(z)dt^2+\frac{dz^2}{z^4(1+b^2z^2)f(z)}+\frac{1+b^2z^2}{z^2}g(z)d\vec{x}^2,
\label{metric}
\end{equation}
where $z$ is the radial coordinate, $\vec{x}\equiv(x_1,x_2,x_3)$ and
\begin{align}
f(z)&=\frac{1+b^2z^2}{z^2}\Gamma^{2\gamma}-\frac{mz^2}{1+b^2z^2}\Gamma^{1-\gamma},\nonumber\\
N(z)&=\Gamma^{-\gamma},\ g(z)=\Gamma^{\gamma},\ \Gamma(z)=1-\frac{b^2z^2}{1+b^2z^2},\ \gamma=\frac{\a^2}{2+\a^2}.
\label{coefficient}
\end{align}
$\alpha$ determines the coupling constant between the gauge field and the dilaton. The constant $b$ is related to $m$, the mass of the black hole and $q$, the charge of the black hole as follows
\begin{equation}
q=\sqrt{\frac{6m}{2+\a^2}}b.
\label{chargemass}
\end{equation}
This background is asymptotically AdS$_5$. The field theory lives on the boundary where is located at $z=0$. The Hawking temperature of the black hole, corresponding to the temperature of the field theory, is
\begin{equation}
T=\frac{b\Gamma(z_h)^{\frac{3\gamma}{2}-1}}{4\pi\sqrt{1-\Gamma(z_h)}}(2(3\gamma-1)-3(2\gamma-2)\Gamma(z_h)),
\label{temperature}
\end{equation}
where $z_h$ is the horizon. Due to the gauge field in the balk, the chemical potential in the field theory is \cite{thermalization}
\begin{equation}
\mu=\frac{b\sqrt{3m}}{\sqrt{2(\a^2+2)}(b^2+\frac{1}{z_h^{2}})}.
\label{chemical}
\end{equation}
In the case of $\alpha=2$, the above background is identical with the one that enjoys a critical point at $\mm=1.1107$ and its phase diagram can be found in \cite{ebrahim}.
To be more precise, using \eqref{temperature} and \eqref{chemical}, one easily obtains
\begin{equation}
bz_h=\frac{1\pm\sqrt{1-\frac{8\mu^2}{\pi^2T^2}}}{\frac{2\mu}{\pi T}},
\label{bhorizon}
\end{equation}
where
\begin{equation}
z_h=\sqrt{\frac{b^2+\sqrt{b^4+4m}}{2m}}.
\label{horizon}
\end{equation}
Therefore, for each value of $\frac{\mu}{T}$, there are two distinct values of $bz_h$ which specify stable and unstable branches of solutions  indicating that there exist a phase transition in field theory and the critical point is where the two branches merge \cite{ebrahim}. The upper (lower) sign in \eqref{bhorizon} correspond to thermodynamically unstable (stable) solutions. For thermodynamically stable solution the Jacobian, $\mathcal{J}=\frac{\partial(s,\rho)}{\partial(T,\mu)}$ is positive \cite{quasi}. $s$ and $\rho$ are entropy and charge density, respectively
\begin{equation}
s\propto\frac{T^{3}(1+b^{2}z_{h}^{2})^{2}}{(2+b^{2}z_{h}^{2})^{3}},
\label{entropy}
\end{equation}
\begin{equation}
\rho\propto \frac{\mu}{T}(2+b^{2}z_{h}^{2})\sqrt{1+b^{2}z_{h}^{2}}.
\label{density}
\end{equation}
Note that with some field redefinition this is the holographic model of the QCD critical point \citep{gubser}. 

Calculating relaxation time and investigation of time evolution of HSC are of interest in this paper. Thus we consider a Vaidya type solution found by replacing $f(z)$ with $f(z,v)$, in which
\begin{equation}
dv=dt-\frac{dz}{z^2\sqrt{(1+b^2z^2)N(z)}f(z)}.
\label{time}
\end{equation}
Note that $t$ and $v$ are equal at the boundary. This replacing has been done with exchanging $m$ and $q$ with $m I(v)$ and $q \sqrt{I(v)}$, respectively. $I(v)$ is the time-dependent function which is chosen
\begin{equation}
I(v)=\frac{1}{2}\left(1+\tanh(\frac{v}{v_0})\right),
\label{tangant}
\end{equation}
where $v_0$ indicates the speed of energy injection into the background. In fact it measures how fast the function can reach a maximum value. In the litrature $v_0 \ll 1$ ($v_0 \gg 1$) is called fast (slow) quench. Different functions which can be chosen instead of \eqref{tangant} have been discussed in \cite{sharifi1,sharifi2} and it seems that the results are independent of the picked functions. By  replacing above equations in \eqref{metric}, the Vaidya AdS (VAdS) like solution is
\begin{equation}
ds^2=-N(z)f(z,v)dv^2-\frac{2}{z^2}\sqrt{\frac{N(z)}{1+b^2z^2}}dvdz+\frac{1+b^2z^2}{z^2}g(z)d\vec{x}^2.
\label{metric2}
\end{equation}
We use the VAdS metric to discuss time evolution of the HSC.
\section{Holographic subregion complexity}
The complexity of the mixed state (a subregion $A$ on the boundary), motivated by the Hubney-Ryu-Takayanagi proposal, is dual to the volume enclosed by the extremal surface $\gamma_{A}$ appearing in the computation of holographic entanglement entropy \citep{alishahiha}, i.e.
\be 
{\cal{C}}_{A}=\frac{V_{\gamma_{A}}}{8\pi R G_{N}},
\label{cv}
\ee
\begin{figure}[H]
\centering
\includegraphics[width=70 mm]{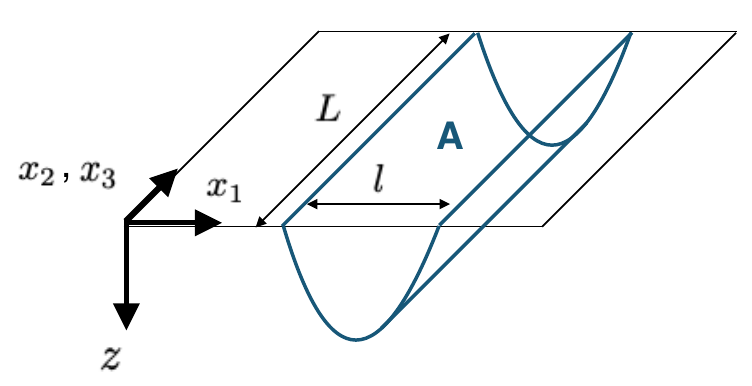} 
\caption{A strip entangling surface of length $l$ and width $L\rightarrow \infty$ in a static AdS background.}
\label{fig0}
\end{figure} 
where $R$ and $G_{N}$ are AdS radius and the Newton constant, respectively. ${\cal{C}}_{A}$ is the HSC for the subregion $A$.
In order to compute $\gamma_A$, at a given time, we consider the subsystem $A$ which is defined as
\begin{align}\label{subsystem}
-\frac{l}{2}<x_1(\equiv x)<\frac{l}{2},\ x_2\in(-\infty,+\infty),\ x_3\in(-\infty,+\infty),
\end{align}
and describes a strip surface of length $l$ and width $L\rightarrow\infty$ as you can see in figure \ref{fig0} in a static AdS background. For a dynamical background $\gamma_{A}$ do not live on a constant time slice. Duo to the symmetry of the strip, the extermal surface $\gamma_A$ can be parametrized as
\be
v=v(x),\ z=z(x),\ v(\frac{l}{2})=t,\ z(\frac{l}{2})=z_0,
\ee
where $z_0$ is a UV cut-off. Then the area of the minimal surface turns out to be
\begin{equation}
S=\frac{L^2}{4G_5}\int_{-l/2}^{l/2} dx \frac{1+b^2z^2}{z^2}g(z)\sqrt{-N(z)f(z,v)v'(x)^2-\frac{2}{z^2}\sqrt{\frac{N(z)}{1+b^2z^2}}z'(x)v'(x)+\frac{1+b^2z^2}{z^2}g(z)}.
\label{surface2}
\ee
Considering the integrand in \eqref{surface2} as a Lagrangian,
where the symmetry of the problem indicates that the turning point of the minimal surface lies at $x_{1}=0$ and we therefore used
\be\label{condition}
v'(0)=z'(0)=0,\ v(0)=v_*,\ z(0)=z_*.
\ee
One can find the equations of motion for $z(x)$ and $v(x)$ and then, using \eqref{condition}, they can be numerically solved to obtain the profiles of $z(x)$ and $v(x)$.
In the case at hand, the volume can be paprametrized by $v=v(x)$ and $z=z(x)$, or equivalently $z=z(v)$. For the background solution \eqref{metric2} the induced metric on the volume is
%The area of the extremal surface using \eqref{metric2} is
\begin{equation}
ds^2=-\left( N(z)f(z,v)+\frac{2}{z^2}\sqrt{\frac{N(z)}{1+b^2z^2}}\frac{\partial{z}}{\partial{v}}\right) dv^2+\frac{1+b^2z^2}{z^2}g(z)\left(dx^2+\sum_{i=1}^{2}dx_i^2\right),
\label{induce}
\end{equation}
and the volume becomes
\begin{equation}
V=2L^2\int_{v_*}^{v(\frac{l}{2})}dv\int_{0}^{x(v)}dx\ \frac{1+b^2z^2}{z^2}g(z)\sqrt{\frac{1+b^2z^2}{z^2}g(z)\left( -N(z)f(z,v)-\frac{2}{z^2}\sqrt{\frac{N(z)}{1+b^2z^2}}\frac{\partial{z}}{\partial{v}}\right)}.
\label{volume2}
\end{equation}
Clearly in the above equation  $\int_{0}^{x(v)}dx$ can be replaced by $x(v)$.
Since the the HSC (or equavalentely volume in the dual gravity) is divergent, we would like to introduce a normalized version of HSC using \eqref{cv}, as follows
\begin{align}
C\equiv\frac{8\pi RG_{N}(\mathcal{C}_{VAdS}-\mathcal{C}_{AdS})}{L^2}=\frac{V-V_{AdS}}{L^{2}},
\label{cv2}
\end{align}
where $\mathcal{C}_{VAdS}$ and $\mathcal{C}_{AdS}$ are the HSC for $A$ in VAdS and AdS geometry respectively. The volumes are defined for the same boundary region such that $V$ in equation \eqref{volume2} reduces to $V_{AdS}$ by setting $b$ and $m$ equal to zero. 

 In oreder to find the relaxation time scale for the complexity, we introduce the following function 
\be
{\epsilon}(t)=|1-\frac{C(t)}{C(\infty)}|.
\ee
Then the relaxation time for HSC, $t_{c}$, is the time at which $\epsilon(t)<10^{-3}$ and stays below this limit forever. We will numerically calculate time scale $t_{c}$ later on.

\section{Numerical results}
In this section we will present our results from the numerical calculation of HSC at $\rm{VAdS}$ metric \eqref{metric2}. We study the evolution of HSC for an out-of-equilibrium process corresponding to $\rm{VAdS}$ background in the gravity dual. 

We are interested in investigating the behavior of HSC as we move towards the critical point, $(\frac{\mu}{T})_*$, in the field theory. In figure \ref{fig1}, we have plotted the final value of the rescaled HSC, $\frac{C_{eq}}{\mu T}$ meaning $\frac{C}{\mu T}$ at $t\rightarrow \infty$, and rescaled relaxation time, $v_{0}^{-1}t_{c}$, as a function of $\frac{\mu}{T}$. In the left panel, $\frac{C_{eq}}{\mu T}$ and $\frac{\mu}{T}$ increase together. By increasing $\frac{\mu}{T}$, the amount of information it takes to specify the final state increases. In the right panel we can observe with increasing $\frac{\mu}{T}$, the time it takes for rescaled HSC to relax decreases. This interesting result means that by increasing $\frac{\mu}{T}$, although the final value of the complexity increases, it takes less time to get complexity equilibrium.

It seems that moving towards the critical point can highlight the difference between two concepts  of complexity. That is, how much information it takes to specify a state and how many operations need to be done to reach the state. Using the terms prevalent in quantum or classical information literatures, we can consider the space of states is described by K classical bits, (010111001...). Our desired task is to start from a state in which all bits are the same and reach the state in which the bits alternate. This target state (010101010...) can be complex from one point of view and simple from another. How many simple operations have to be done to reach the target state, or how long it has to run, do depend on the number of bits and thus (010101010...) is complex  from this aspect. While if we consider how much information is needed to specify the target state it is simple since  it does not depend on the number of bits \cite{susskind}. It seems that we observe such behavior in the case that we study. On the one hand complexity increases due to the amount of information required to specify the final state by increasing $\frac{\mu}{T}$ and on the other hand, based on the time it takes to run, the final state or the task that has to be done for getting to the final state is simple. We speculate this decrease of the time it takes to reach the final state or the simplicity of the final state from this point of view could be because of movement to the critical point of the theory, since near the critical point universality classes are characterized with a large scale behavior and the theory can be described with fewer parameters.

\begin{figure}[H]
\centering
\includegraphics[width=70 mm]{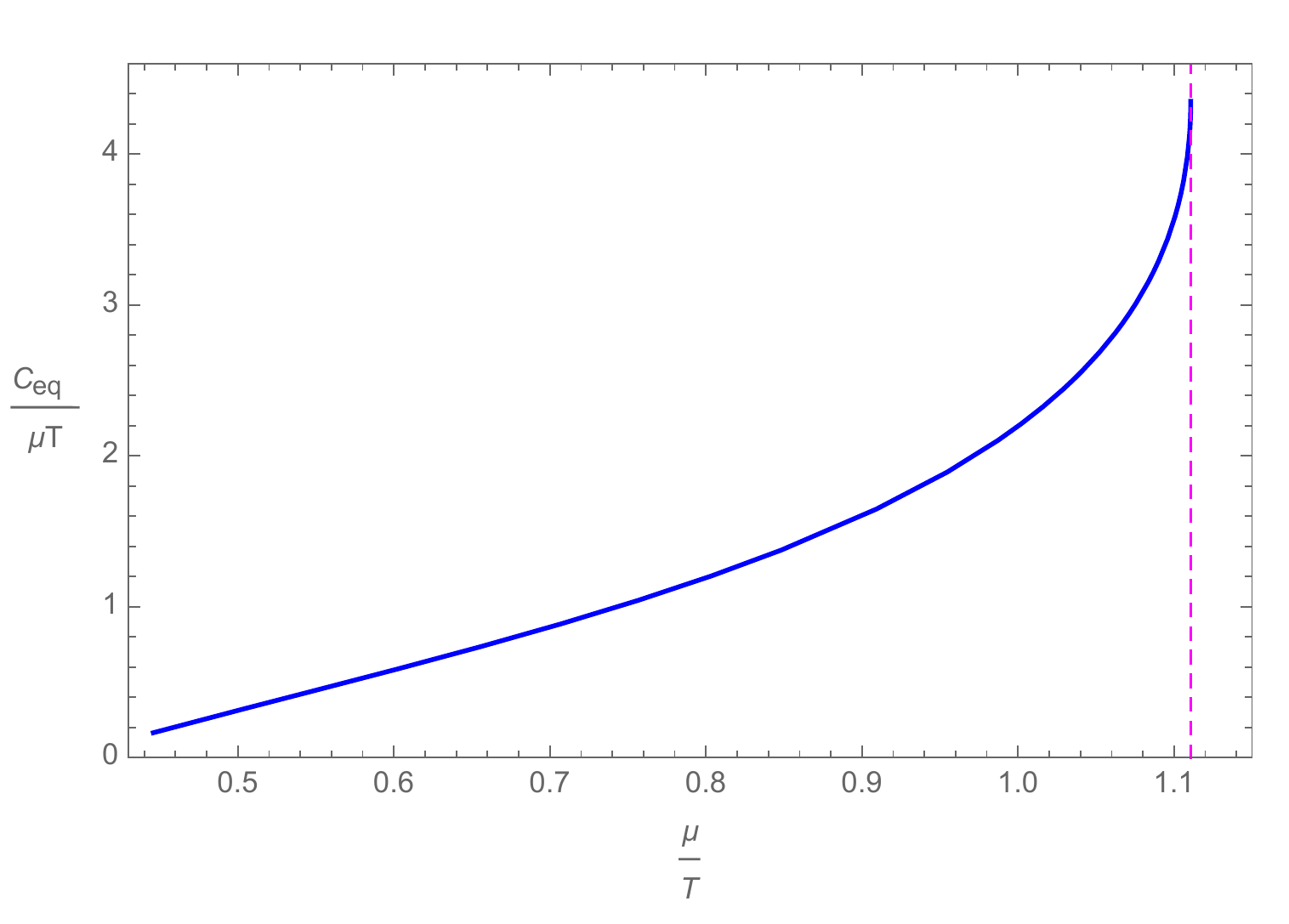}
\includegraphics[width=70 mm]{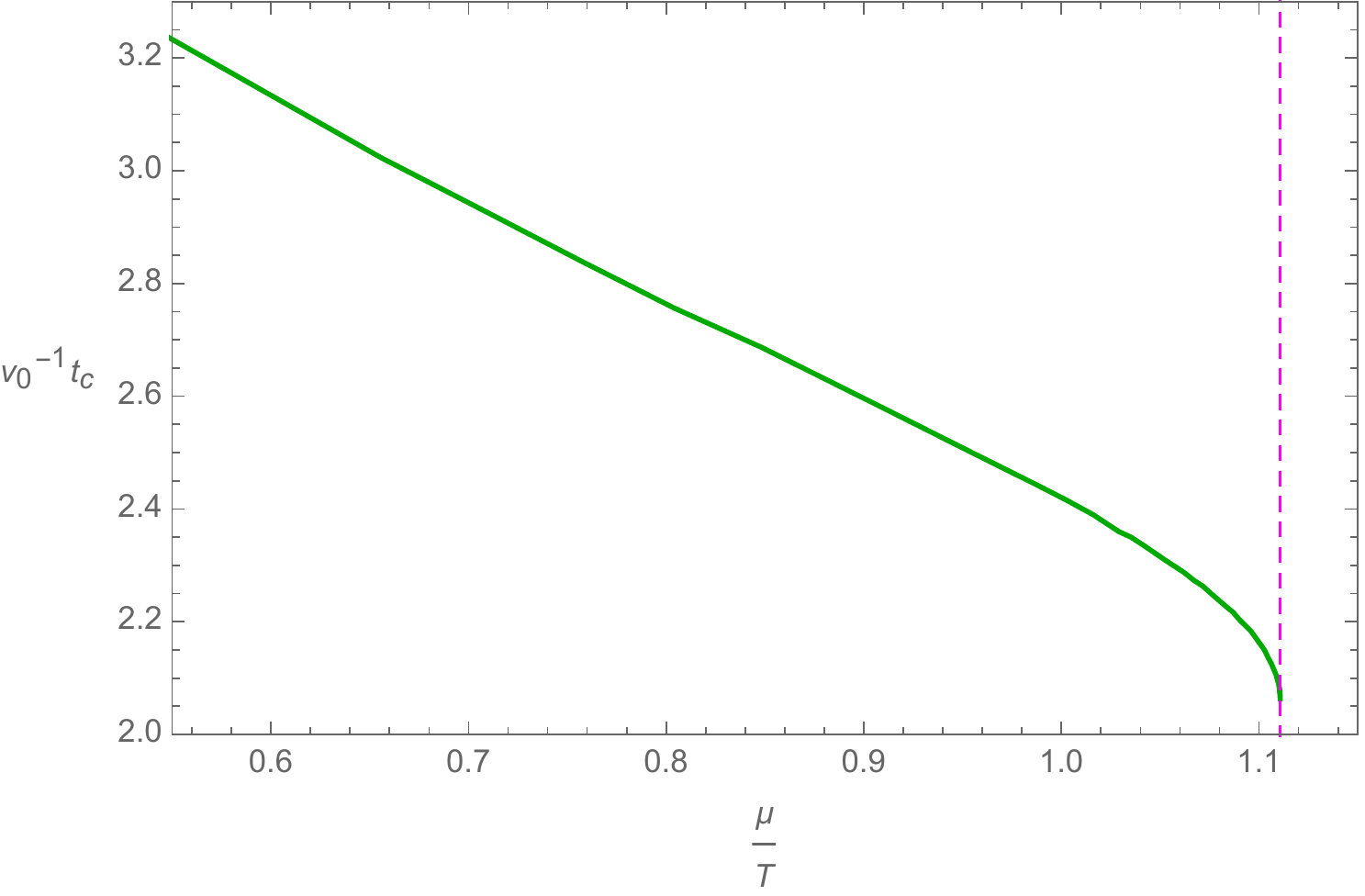}   
\caption{Left: The behavior of the final value of the rescaled HSC, $\frac{C}{\mu T}|_{t\rightarrow\infty}=\frac{C_{eq}}{\mu T}$, as a function of $\frac{\mu}{T}$ for $l=0.3$ and $v_{0}=3$. The magenta dashed line indicates the critical point which is at $(\frac{\mu}{T})_*$. Right:  The behavior of the rescaled relaxation time, $v_{0}^{-1}t_{c}$, with respect to $\frac{\mu}{T}$.}
\label{fig1}
\end{figure} 

An interesting question is whether HSC understands about the phase structure of the theory. As you can see in figure \ref{fig1}, the slope of $\frac{C_{eq}}{\mu T}$  approaches infinity at $(\frac{\mu}{T})_*$. We can define 
\begin{align}
&\frac{d\left(\frac{C_{eq}}{\mu T}\right)}{d\frac{\mu}{T}}(i)=\frac{\frac{C_{eq}}{\mu T}(i+1)-\frac{C_{eq}}{\mu T}(i)}{\frac{\mu}{T}(i+1)-\frac{\mu}{T}(i)},
\label{slope1}
\end{align}
where $i$ indicates the $i$th point of data points.
The slope of the curve for the points near the critical point can be fitted with a function of the form $(\frac{\pi}{2\sqrt{2}}-\frac{\mu}{T})^{-\theta}$ where $\theta$ is the dynamical critical exponent \cite{ebrahim}. In figure \ref{fig2}, we have plotted the slope \eqref{slope1}, near the critical point. We get $\theta=0.50166$ and $\theta=0.513942$ for slow and fast quench respectively, which are in perfect agreement with one is obtained from the behavior of scalar quasi-normal modes \cite{quasi}. Hence HSC is a good observable for probing the dynamical critical exponent. Various observables have been studied to find critical exponent \citep{ebrahim2,amrahi,hajar}.

\begin{figure}[H]
\centering 
\includegraphics[width=70 mm]{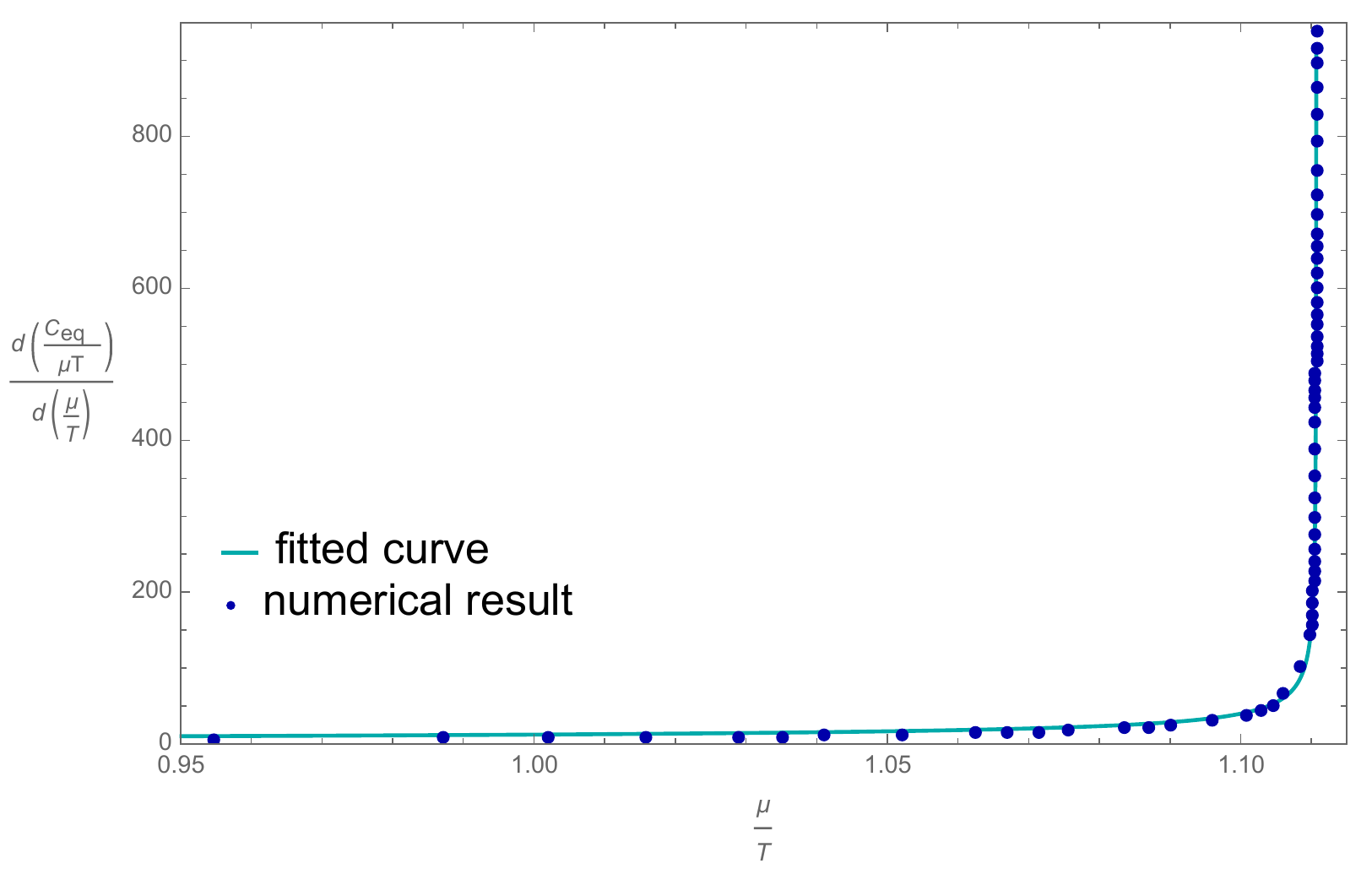}   
\includegraphics[width=70 mm]{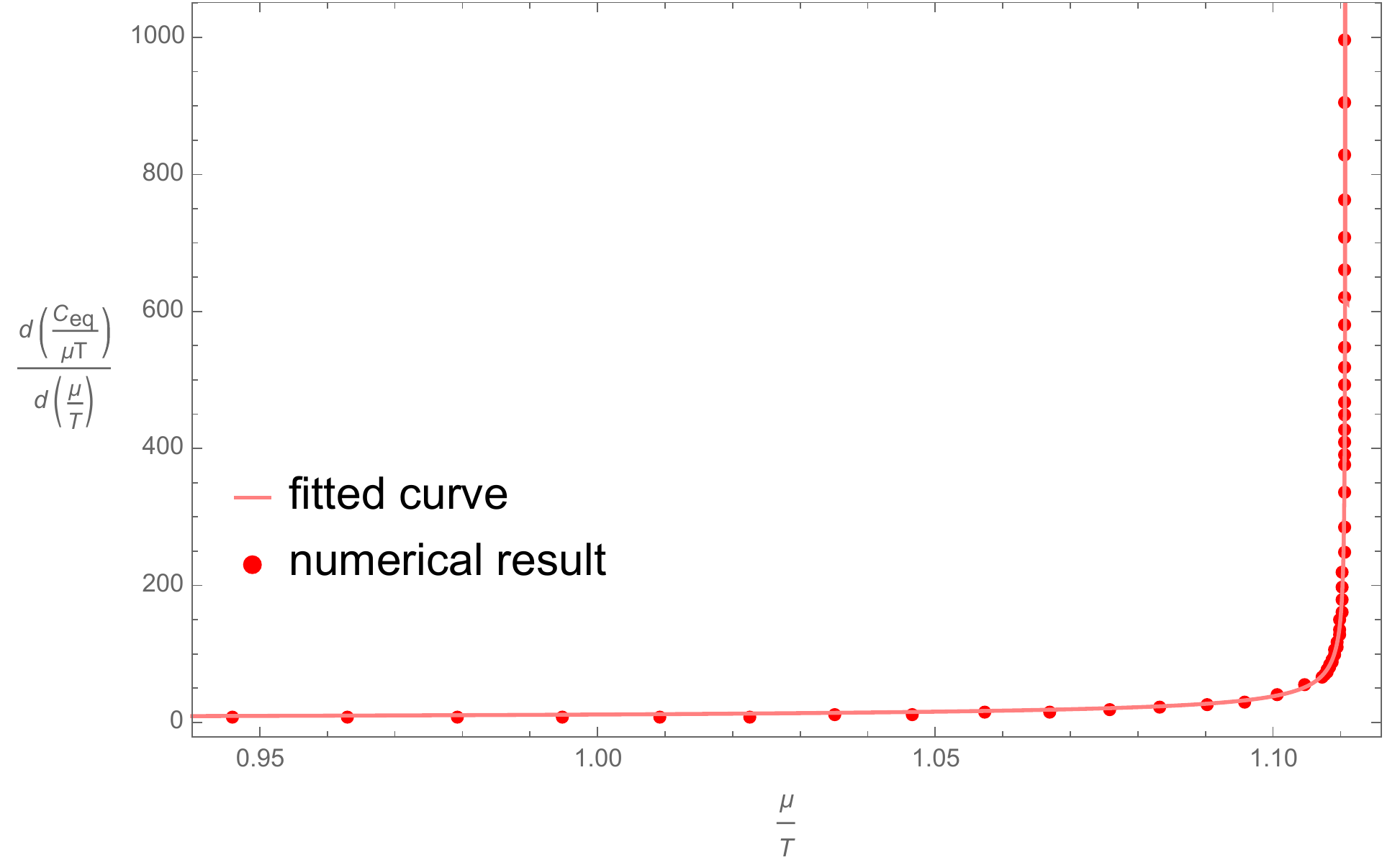}
\caption{Left: The slope of the final value of the rescaled HSC, $\frac{C_{eq}}{\mu T}$ as a function of $\frac{\mu}{T}$ for $l=0.3$ and $v_{0}=3$, near the critical point. The light blue curve is the function, $\left(\frac{\pi}{2\sqrt{2}}-\frac{\mu}{T}\right)^{-\theta}$, fitted with the data by $\theta=0.501662$. Right: The slope of $\frac{C_{eq}}{\mu T}$ as a function of $\frac{\mu}{T}$ for $l=0.3$ and $v_{0}=0.05$. The pink curve fitted with the data by $\theta=0.513942$.}
\label{fig2}
\end{figure} 

\begin{figure}[H]
\centering
\includegraphics[width=70 mm]{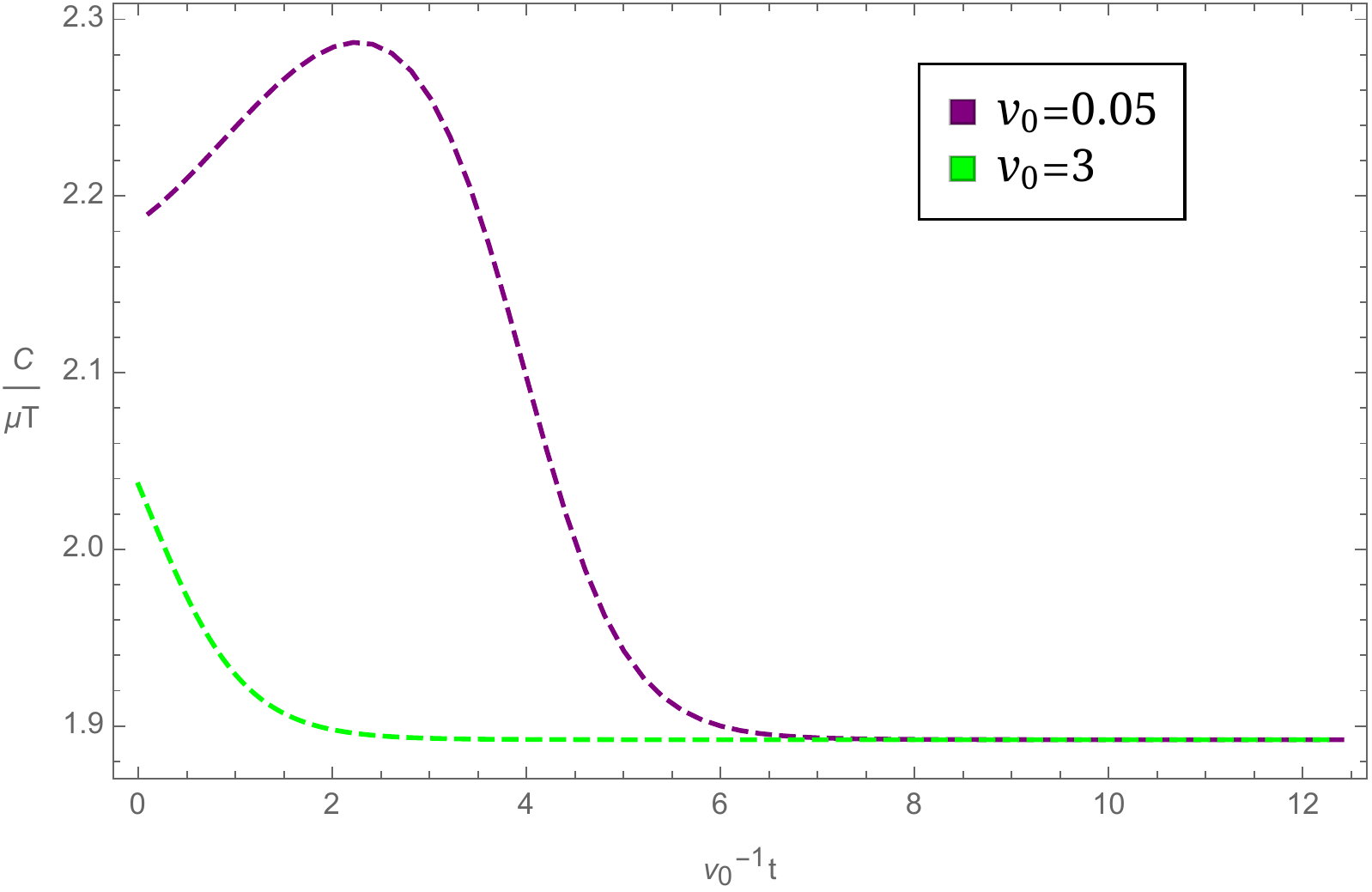}
\includegraphics[width=70 mm]{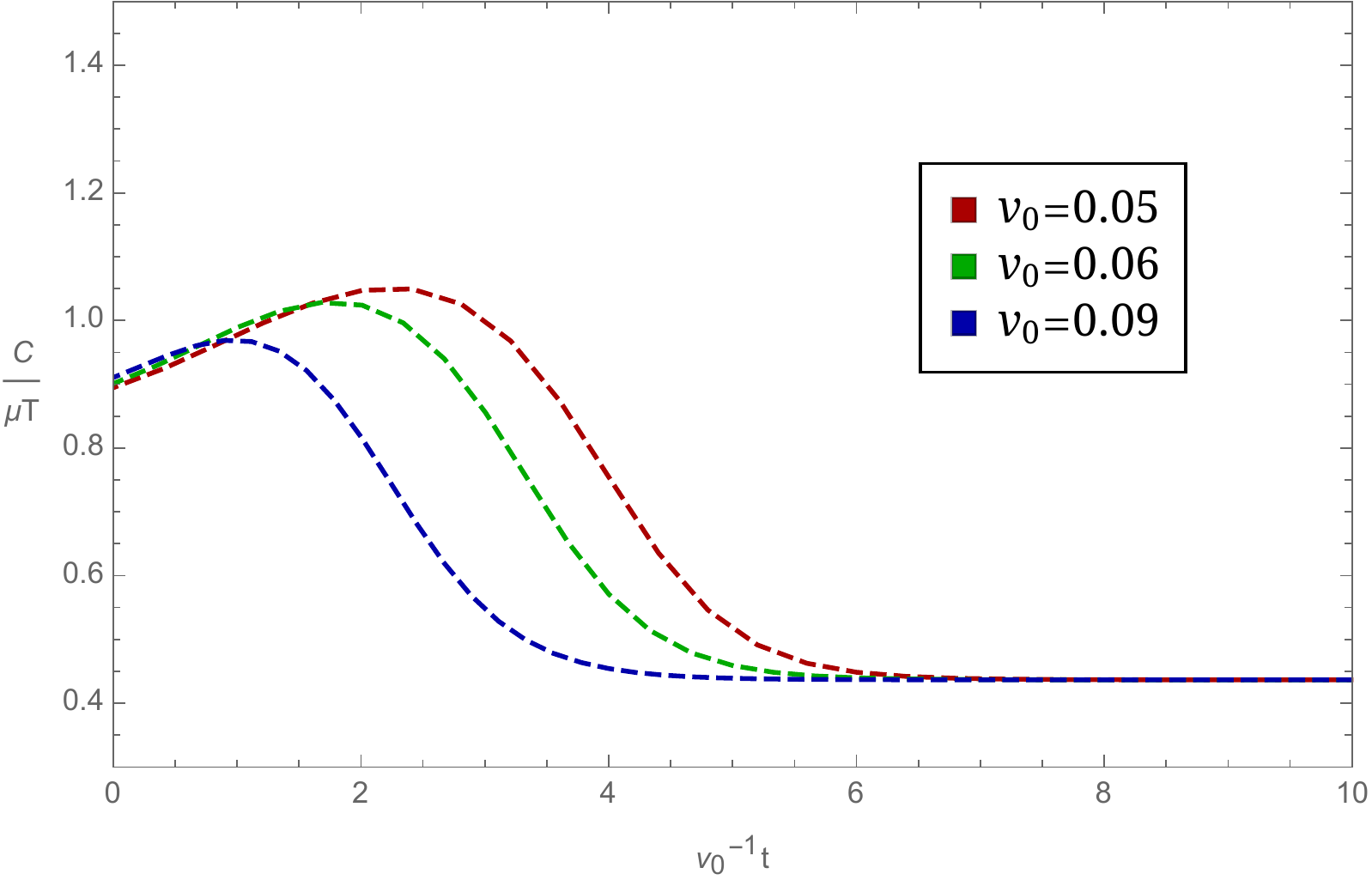}   
\caption{Left: The rescaled HSC, $\frac{C}{\mu T}$ as a function of $v_{0}^{-1}t$ for $l=0.3$ and two values of $v_{0}$ at fixed $\frac{\mu}{T}=0.954716$ . Right: The rescaled HSC in terms of rescaled time for three values of $v_{0}$ at fixed $\frac{\mu}{T}=0.546466$ for $l=0.3$.}
\label{fig3}
\end{figure} 
In figure \ref{fig3}, left, we have plotted rescaled HSC as a function of rescaled time for two different values of $v_{0}$. At slow quench, $v_{0}=3$, the system has enough time to equilibrate during the transition and undergoes an adiabatic evolution. However, at fast quench, $v_{0}=0.05$, with a rapid altering, the rescaled HSC experiences an increase at the early stage that reaches a maximum value and then decreases to a constant value at late time, $C_{eq}$. The faster the energy injection, the sooner the quench happens, and the earlier the system reaches complexity equilibrium. Different manner of energy injection into the system leads to different responses during early time interval. Therefore, if we know only about the $T$ and $\mu$ in the boundary field theory, the rescaled HSC can distinguish between fast or slow quench.
To study the effects of $v_{0}$ on evolution of rescaled HSC, we have plotted rescaled HSC as a function of rescaled time in the right panel of figure \ref{fig3}. With the increase of $v_{0}$, in the range of fast quench, the maximum value which rescaled HSC can reach becomes smaller and is achieved sooner. 

Due to the second law of quantum complexity,  a process with decreasing complexity is counterintuitive in the boundary field theory from the thermodynamical point of view \cite{uncom2}. This behavior has been reported in \citep{quench,quench2,quench3,quench4} and an attempt has been made to interpret it qualitatively using interpretation of HSC as a purification complexity \cite{quench4}. What we want to emphasize are as follows: 
 \begin{itemize}
\item As shown in figure \ref{fig3}, left and right,  in the evolution from the initial state corresponding to the VAdS background in initial time at zero temperature, to the thermal state corresponding to the VAdS background in final time, complexity decreases. Initial rescaled HSC, $C_{0}$ is always bigger than final rescaled HSC, $C_{eq}$. According to quantum statistical mechanics, there is an ensemble of microstates corresponding to a given mixed macrostate. In time evolution of the mixed state under study, at $t=0$ corresponding to the zero of entropy, there exists one microstate, which is called {\it{unique configuration}}, but nearing the equilibrium with increasing temperature, the number of these microstates increases. In order to specify the macrostate, it is not necessary to know the details of the underlying system. As in statistical systems, for instance, the speed of sound is determined only with the macroscopic characteristics of the system, i.e. pressure and density, in our case to give an effective description of the mixed macrostate we do not need the details of the microstates that lead to it. Therefore, we specify the state with less information than the zero temperature case in initial time which corresponded to one microstate.
%and consequently our information about the state decreases. In other words, when we do not have exact knowledge of the microstates of a system, the system is said to be in a mixed state. By turning on the external source or increasing temperature to reach the equilibrium, the lack of this knowledge increases due to the increasing number of microstates which lead to the mixed macrostate. Thus,  the information which is needed for specifying this mixed state decreases with time.
Since the underlying field theory is conformal, our results are $\frac{\mu}{T}$ dependent i.e. increasing chemical potential in this case is like decreasing temperature and vice versa. Hence, this explanation is in agreement with the left panel of figure \ref{fig1}.

\item Decreasing complexity from its maximum possible value or lack of complexity is a resource for a computational work \cite{uncom1,uncom2}. In analogy with thermodynamic free energy that is used to do a work, uncomplexity is a resource for doing a directed computation \cite{uncom2}. Therefore a quantum computer which has been run and has reached maximum complexity is useless for doing any computations. With the decrease of complexity of the black hole as a computational machine, we have a resource for performing directed quantum computation. We can interpret relative complexity of the initial state and the final state as a resource which is expended to reach the final state. We call it as follows 
 \be 
\Delta C=C_{eq}-C_{0}.
\label{deltac}
\ee
In figure \ref{fig4}, left, rescaled HSC as a function of rescaled time for three values of $\frac{\mu}{T}$ have been plotted. As we near the critical point, $C_{0}$ and $C_{eq}$ increase and $C_{eq}$ is smaller than $C_{0}$. In the middle and right panel of figure \ref{fig4}, $\Delta C$ as a function of $\frac{\mu}{T}$ have been plotted for fast and slow quench. Nearing the critical point, $\Delta C$ decreases and at the critical point, the magenta dashed line, the slope of the curve approaches infinity. It seems that with the decrease of the needed resource for reaching the final state, $\Delta C$, as it approaches the critical point, the time it takes to expend the resource decreases too, in agreement with the right panel of figure \ref{fig1}.

\item The system we study consists of a state and its environment. In this particular case, the external source that injects energy into the boundary field theory is the environment. In initial time the external source is zero and thus its complexity or the information for specifying it is zero too. It is obvious during the evolution when the external source value becomes non-zero, we need the information to specify it. Therefore, in this evolution, complexity of external source increases. In contrast, the effect of the external source on the state as you can see in left panel in figure \ref{fig4}, is the decrease of complexity. We speculate that in this process, complexity of the environment increases and complexity of the state decreases. Therefore, it seems the second law of complexity is not violated for the whole system.
\end{itemize}
\begin{figure}[H]
\centering 
\includegraphics[width=58 mm]{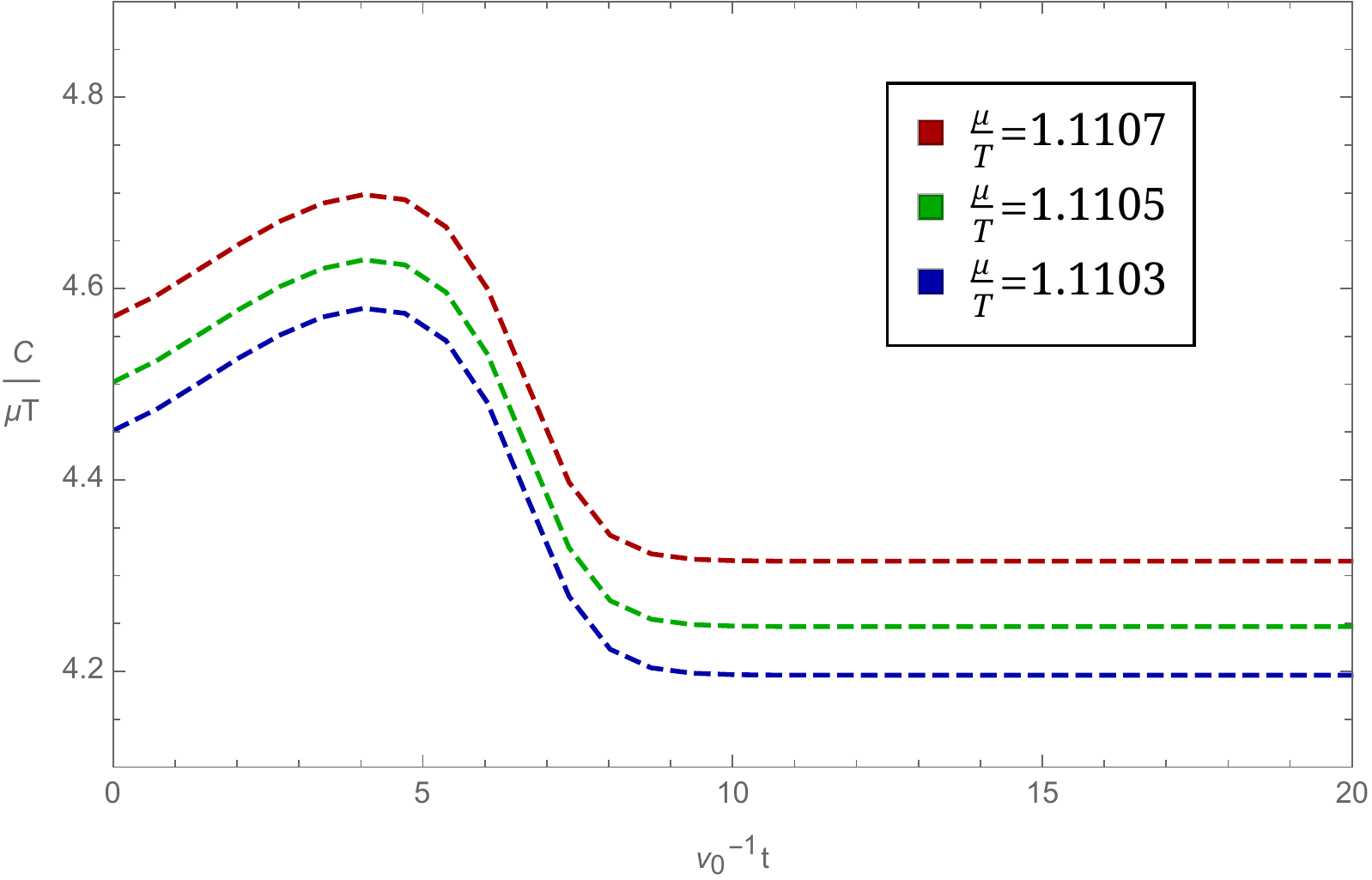}
\includegraphics[width=58 mm]{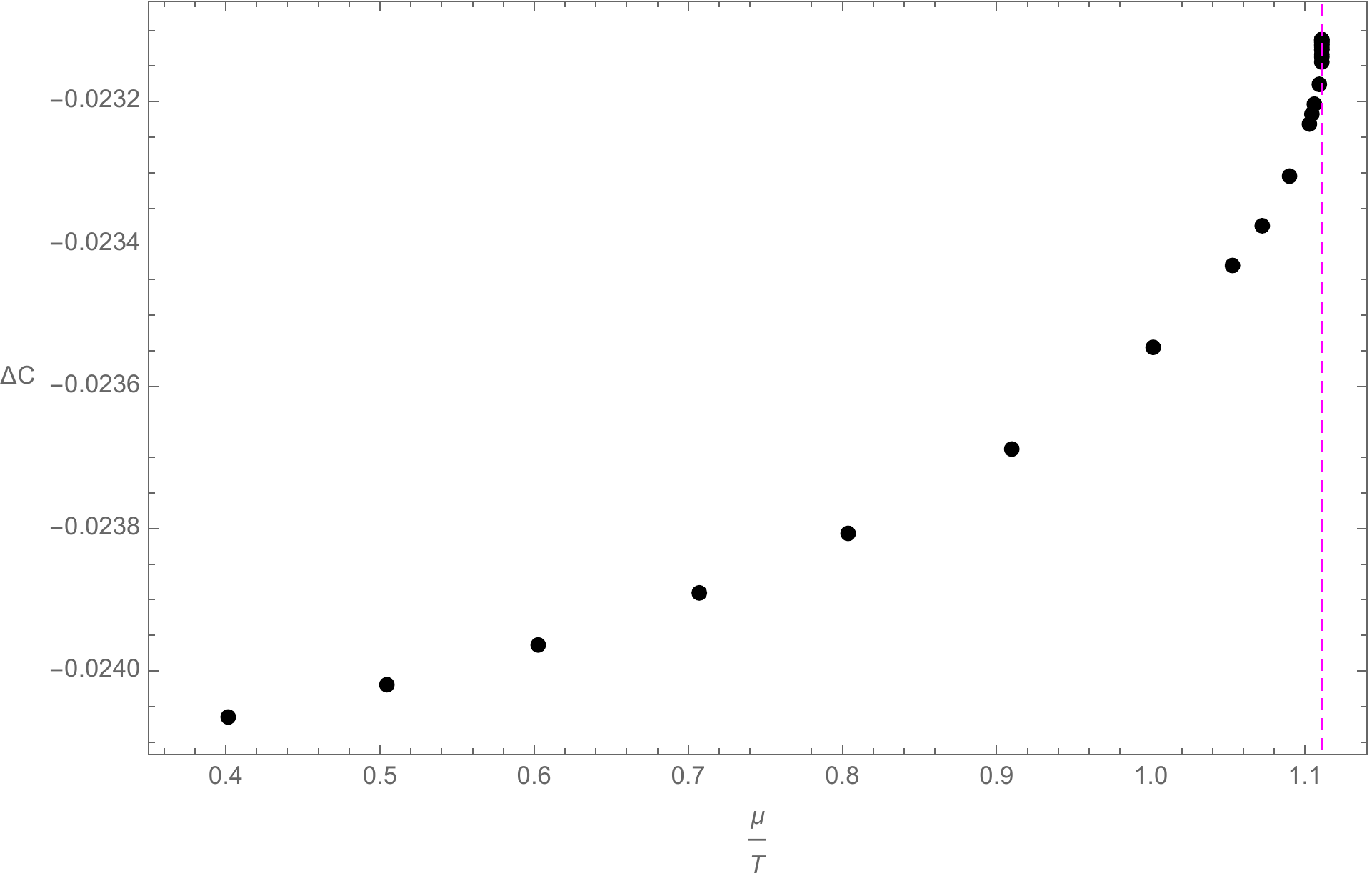}
\includegraphics[width=58 mm]{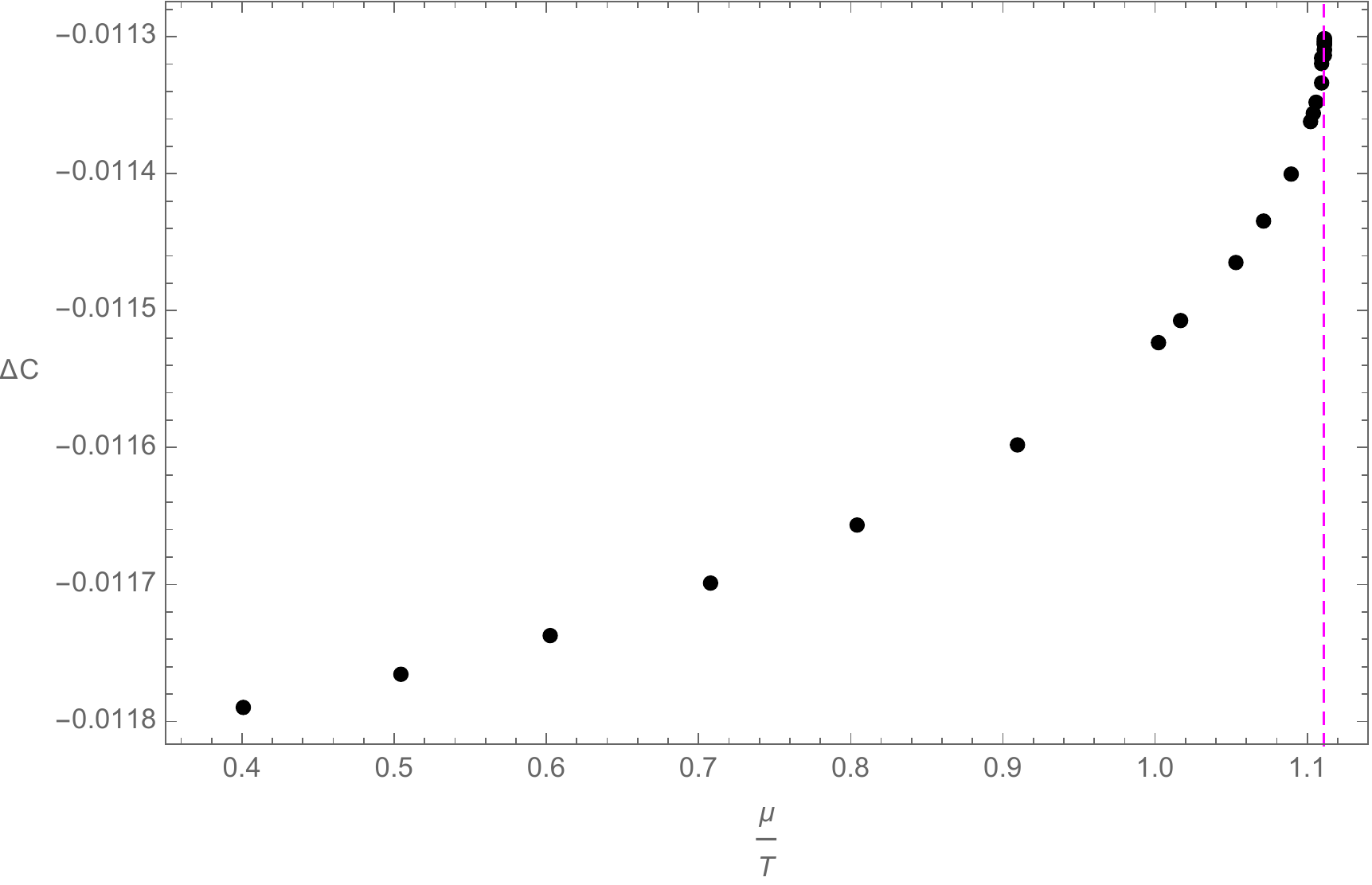}   
\caption{Left: The rescaled HSC, $\frac{C}{\mu T}$ as a function of $v_{0}^{-1}t$ for three different values of $\frac{\mu}{T}$, $l=0.3$ and $v_{0}=0.05$. Middle: $\Delta C$ in terms of $\frac{\mu}{T}$ for $v_{0}=0.05$ and $l=0.3$. The magenta dashed line shows the critical point. Right: $\Delta C$ in terms of $\frac{\mu}{T}$ for $v_{0}=3$ and $l=0.3$. }
\label{fig4}
\end{figure} 

%Before ending this section, we would like to note that it is expected the Lloyd's bound apply when subregion complexity coincides with the complexity of the pure state \citep{time}. Since there is a limitation in our numerical method for large $l$, an analytical method like \cite{quench2} can be used to investigate this limit.

\end{document}